\documentclass{jpsj-suppl}
\usepackage{txfonts} 
\usepackage{wrapfig}
\title{1+1 Large $N_c$ QCD and its Holographic Dual \\ 
$\sim$ Soliton Picture of Baryons in Single-Flavor World}

\author{Hideo \textsc{Suganuma}$^{1}$, Yuya \textsc{Nakagawa}$^{2}$, and Kohei \textsc{Matsumoto}$^{2}$
}

\inst{$^{1}$Department of Physics, Kyoto University, Kitashirakawaoiwake, Sakyo, Kyoto 606-8502, Japan \\
$^{2}$Yukawa Institute for Theoretical Physics (YITP), Kyoto University, Sakyo, Kyoto 606-8502, Japan}

\email{suganuma@scphys.kyoto-u.ac.jp}

\recdate{October 6, 2016}

\abst{
We study baryons in holographic QCD corresponding to 1+1 dimensional 
single-flavor ($N_f$=1) QCD for the first time. 
We formulate 1+1 QCD using an $S^1$-compactified D2/D8/$\overline{\rm D8}$ branes 
in the superstring theory, and describe the baryon 
as a topological configuration in 1+1 $N_f$=1 QCD, 
corresponding to $\Pi_1({\rm U(1)})={\bf Z}$. 
Unlike 1+3 QCD with $N_f \ge 2$, however, 
we find that the low-dimensional baryonic soliton is 
generally unstable against a scale transformation/variation and swells infinitely 
in 1+1 $N_f$=1 QCD at the leading of large $N_c$. 
We thus point out a serious difficulty on the soliton picture of baryons in large $N_c$ 
in the single-flavor world in both 1+1 and 1+3 QCD.
We also compare the low-dimensional holographic baryon 
with the Abrikosov vortex, i.e., a stable topological configuration 
in Type-II superconductors.}

\kword{holographic QCD, low-dimensional QCD, baryon, soliton}

\begin{document}
\maketitle

\section{Introduction: Holographic QCD and Baryons in Large $N_c$}

Since 1973, quantum chromodynamics (QCD) has been established as 
the fundamental theory of the strong interaction. 
Nevertheless, it is very difficult to solve QCD directly in an analytical manner, 
and many effective models of QCD have been used instead of QCD, 
but most models cannot be derived from QCD and 
its connection to QCD is unclear.
To analyze nonperturbative QCD, 
the lattice QCD Monte Carlo simulation has been also used 
as a first-principle calculation of the strong interaction. 
However, it has several weak points. 
For example, the state information (e.g. the wave function) is severely limited, 
because lattice QCD is based on the path-integral formalism. 
Also, it is difficult to take the chiral limit, because 
zero-mass pions require infinite volume lattices.
There appears a notorious ``sign problem'' at finite density.

On the other hand, holographic QCD \cite{W98,SS05,NSK07} 
has a direct connection to QCD, 
and can be derived from QCD in some limit. 
In fact, holographic QCD is equivalent to infrared QCD 
in large $N_c$ and strong 't~Hooft coupling $\lambda$,
via gauge/gravity correspondence.
Remarkably, holographic QCD is successful to reproduce many 
hadron phenomenology such as vector meson dominance, 
the KSRF relation, hidden local symmetry, 
the GSW model and the Skyrme soliton picture \cite{SS05}.
Unlike lattice QCD simulations, 
holographic QCD is usually formulated in the chiral limit, 
and does not have the sign problem at finite density \cite{NSK07}. 


In general, when large $N_c$ is taken, 
QCD reduces a weakly interacting theory of mesons (and glueballs), 
and the baryon is described as a Skyrmion, 
i.e., a topological chiral soliton of mesons 
(mainly Nambu-Goldstone bosons) \cite{W79}
Actually, in holographic QCD with large $N_c$, 
the theory is described by pseudoscalar, vector and axial-vector mesons \cite{SS05}, 
and baryons do not appear as explicit degrees of freedom but 
appear as spatially-extended topological solitons composed of mesons \cite{NSK07}. 

\section{Puzzle in Single Flavor ($N_f$=1) World in Large $N_c$}

In large $N_c$ QCD, the effective theory includes only meson degrees of freedom,
and therefore one has to take the soliton picture for the description of baryons \cite{W79}.
In our real world with $N_f \ge 2$, 
there occurs spontaneous breaking of the chiral symmetry, i.e., 
${\rm SU}(N_f)_L \times {\rm SU}(N_f)_R \rightarrow {\rm SU}(N_f)_V$, 
and baryons can be described as topological chiral solitons, 
according to the nontrivial homotopy group $\Pi_3({\rm SU}(N_f)_A)={\bf Z}$ \cite{W79}.
Actually, this topological chiral-soliton picture of baryons is successful 
for the semi-quantitative description of baryons \cite{ANW83}.

In the single-flavor ($N_f$=1) world, however, baryons cannot be 
described as topological objects, because of absence of the topological charge, 
i.e., $\Pi_3({\rm U(1)}_A)=1$.
Of course, QCD with single flavor ($N_f$=1) is a possible quantum field theory. 
Actually, if the Higgs coupling to d, s-quarks were large enough, 
the single-flavor world would be realized.
In $N_f$=1 QCD, there appear only 
a massive pseudoscalar meson $\eta'(\bar u \gamma_5u)$, 
a vector meson $\omega(\bar u \gamma^\mu u)$, 
and a baryon $\Delta^{++}(uuu)$, as low-lying hadrons. 
Unlike the $N_f \ge 2$ case, 
the single-flavor QCD does not have the topological charge 
because of $\Pi_3({\rm U(1)}_A)$=1, 
and therefore baryons cannot be described as topological objects.
Thus, in the single-flavor world, 
it is difficult to describe baryons with mesons in large $N_c$, 
where baryons do not appear explicitly.
This is an open problem still now.

In this context, we notice that 1+1 single-flavor QCD 
has the topological object, corresponding to the nontrivial homotopy group 
$\Pi_1({\rm U(1)})={\bf Z}$. 
In fact, as a natural possibility, it is expected that the baryon can be described 
as the topological object in 1+1 single-flavor QCD, 
like 1+3 QCD with $N_f \ge 2$. 
This is a motivation to investigate baryons in 1+1 single-flavor QCD, 
especially in large $N_c$, where baryons do not appear explicitly.

\section{Holographic QCD corresponding to 1+1 Single-Flavor QCD and Baryons}

The baryons in 1+1 QCD have been usually studied with 
the bosonization technique \cite{FS93}. However, 
to investigate the topological aspect relating to $\Pi_1({\rm U(1)})={\bf Z}$, 
one has to describe baryons as topological objects. 
For this purpose, holographic QCD is suitable, because 
baryons appear as topological objects in this framework \cite{HSSY07}.

\begin{figure}
\vspace{-5mm}
\begin{center}
\includegraphics[width=100mm,clip]{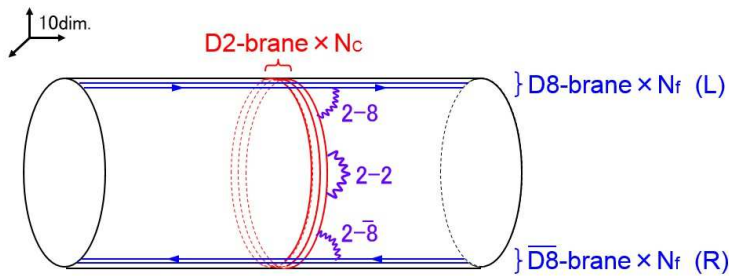}
\end{center}
\vspace{3mm}
\caption{The $S^1$-compactified D2/D8/$\overline{\rm D8}$-brane system 
which gives 1+1 QCD with $N_c$ color and $N_f$ flavor.}
\label{HQCD}
\end{figure}

Similarly to the Sakai-Sugimoto model,
massless 1+1 QCD can be constructed with 
an $S^1$-compactified D2/D8/$\overline{\rm D8}$-brane system 
\cite{YZ11}, 
as shown in Fig.~\ref{HQCD}.
Here, $N_c$ D2-branes give color degrees of freedom, 
and $N_f$=1 D8-brane gives flavor degrees of freedom. 
The gluons appear as 2-2 string modes on $N_c$ D2-branes, and 
the left/right quarks appear as 2-8 string modes 
at the cross point between 
D2 and D8/$\overline{\rm D8}$ branes. 
In this paper, we use the $M_{KK}=1$ unit \cite{SS05}. 

In large $N_c$, $N_c$ D2-branes are extremely massive and can be replaced by 
a gravitational background, via the gauge/gravity correspondence. 
From this D2/D8/$\overline{\rm D8}$-brane system, 
the effective theory of massless 1+1 QCD is derived as 
a 1+2 dimensional U($N_f$=1) gauge theory in the flavor space,
\begin{eqnarray}
S=\frac{N_c}{8\pi}\int dt dx dz \left[ f(z)F_{\mu z}F^{\mu z}
-g(z)\frac12F_{\mu\nu}F^{\mu\nu}+\epsilon^{LMN} A_LF_{MN} \right],
\label{HQCDaction}
\end{eqnarray}
at the leading of $1/N_c$ and $1/\lambda$ expansion \cite{YZ11}.
Remarkably, this theory can be treated at the classical level. 
Here, $z$ is the extra spatial dimension appearing in holographic QCD, 
and $f(z)\equiv (1+z^2)^{1/2}$ and $g(z) \equiv(1+z^2)^{-11/10}$ appear as 
the gravitational effect from $N_c$ D2-branes \cite{YZ11}.

The action (\ref{HQCDaction}) has two parts, i.e., 
the Dirac-Born-Infeld (DBI) action (the 1st and 2nd terms) and 
the Chern-Simons (CS) three-form (the last term).
[The Greek index runs over (0,1)= ($t$, $x$) and 
the capital index runs over (0,1,2)= ($t$, $x$, $z$).]
In 1+1 QCD, the CS term is the leading order as well as the DBI terms \cite{YZ11}, 
while the CS term is subleading of $1/\lambda$ expansion 
compared with the leading DBI term in 1+3 QCD \cite{NSK07}.

In this two-dimensional spatial system on ($x$, $z$), 
we note a topological charge, called the Pontryagin index, 
\begin{eqnarray}
Q \equiv \int dx dz \frac{1}{4\pi}\epsilon_{ij}F_{ij}
=\frac{1}{2\pi} \int dx dz F_{xz}=\frac{1}{2\pi} \int dx dz H \in {\bf Z},
\label{Pont}
\end{eqnarray}
which is an integer, according to $\Pi_1$(U($N_f$=1))={\bf Z}. 
Here, $H \equiv F_{xz}$ is the magnetic field in the U($N_f$=1) gauge theory, 
and $Q$ is the total magnetic flux divided by 2$\pi$.
%
From the holographic viewpoint, 
this topological charge corresponds to the baryon number, 
similarly in the holographic dual of 1+3 QCD \cite{HSSY07}. 
On the topological object in the two-dimensional space ($x$, $z$), 
we note that its direct analogue is 
the Abrikosov vortex with ``quantization of the magnetic flux'' 
in Type-II superconductors.

Next, we consider the topological soliton solution 
corresponding to the (multi)baryon with the baryon number $B=Q (\in {\bf Z})$.
For the calculation, we take the temporal gauge $A_0=0$, 
which leads to the ordinary canonical formalism. 
%
From action (\ref{HQCDaction}), the Lagrangian density reads 
\begin{eqnarray}
{\cal L}=\frac{N_c}{8\pi}\left[f(z)\{ \dot A_z^2-(\partial_x A_z-\partial_z A_x)^2\}
+g(z) \dot A_x^2 +2(A_z \dot A_x-A_x \dot A_z) \right], 
\end{eqnarray}
and the Hamiltonian density 
${\cal H} \equiv \Pi_{A_x}\dot A_x + \Pi_{A_z}\dot A_z -{\cal L}$
with $\Pi_{A_i} \equiv \delta L/\delta \dot A_i$ 
is written as 
\begin{eqnarray}
{\cal H}=\frac{N_c}{8\pi}\left\{
g(z){\dot A}_x^2+f(z){\dot A}_z^2+f(z)(\partial_x A_z-\partial_z A_x)^2
\right\} 
=\frac{N_c}{8\pi}\left\{g(z) E_x^2+f(z) E_z^2+f(z) H^2
\right\},
\end{eqnarray}
where the CS term disappears and $E_i=\dot A_i$ in the temporal gauge. 
Here, ${\cal H}$ is non-negative because of $f(z) > 0$ and $g(z) > 0$.
Then, the total energy $M[\vec A]$ 
of the configuration $\vec A=(A_x, A_z)$ is given by 
\begin{eqnarray}
M[\vec A]=
\frac{N_c}{8\pi}\int dx dz \left\{g(z) E_x^2+f(z) E_z^2+f(z) H^2
\right\}.
\end{eqnarray}

We now consider the ground-state soliton, i.e., the lowest (multi)baryon state, 
under the topological constraint of 
$B=Q \equiv \frac{1}{2\pi}\int dx dz H (\in {\bf Z})$.
Since this topological condition does not act on the electric field $\vec E$,
we can take $E_x=E_z=0$ for the ground-state soliton without change of 
the topological charge $Q$, so that $\vec A$ is $t$-independent.
Thus, the ground-state soliton mass is described as 
\begin{eqnarray}
M[\vec A(x,z)] =\frac{N_c}{8\pi}\int dx dz f(z) H^2
=\frac{N_c}{8\pi}\int dx dz f(z) \left[\partial_x A_z(x,z)-\partial_z A_x(x,z)\right]^2.
\label{Bmass}
\end{eqnarray}

\section{A Scale Instability of Holographic Baryons in 1+1 Single-Flavor QCD}

From Eq.(\ref{Bmass}), one can investigate the (multi)baryonic solutions 
in holographic QCD corresponding to 1+1 QCD with $N_f=1$, 
at the leading of $1/N_c$ and $1/\lambda$ expansions.
As a remarkable conclusion, 
we find that all the (multi)baryonic solutions are generally {\it unstable} 
against some scale transformation (scale variation) and {\it swell infinitely}.
We show this scale instability of baryonic solitons below.

Suppose we obtain a topological (baryonic) solution 
$\vec A^{\rm sol}(x,z)=(A_x^{\rm sol}(x,z), A_z^{\rm sol}(x,z))$, 
which minimizes the mass $M[\vec A(x,z)]$ 
and satisfies the topological condition of $B=Q[\vec A(x,z)](\in {\bf Z})$.
As a general property of the solution, its total energy $M$ 
must be a minimum, i.e., 
\begin{eqnarray}
M[\vec A^{\rm sol}] \le M[\vec A^{\rm sol}+\delta \vec A], 
\end{eqnarray}
against any small variations $\delta \vec A(x,z)$ 
consistent with the topological condition 
$B=Q[\vec A^{\rm sol}+\delta \vec A](\in {\bf Z})$. 

\begin{figure}
\vspace{-5mm}
\begin{center}
\includegraphics[width=120mm,clip]{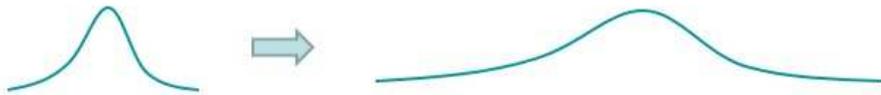}
\end{center}
\caption{The scale instability of the baryonic soliton configuration 
in the holographic dual of 1+1 QCD with $N_f$=1. 
All the (multi)baryonic solutions swell infinitely 
at the leading of $1/N_c$ and $1/\lambda$ expansions.}
\label{scale}
\end{figure}

As a simple variation, we consider a ``scaled configuration'' of 
\begin{eqnarray}
\vec A_\lambda(x,z) 
\equiv (\lambda A_x^{\rm sol}(\lambda x, z), A_z^{\rm sol}(\lambda x, z)),
\end{eqnarray}
which includes the original solution $\vec A^{\rm sol}(x,z)$ at $\lambda=1$, i.e., 
$
\vec A_{\lambda=1}(x,z)=\vec A^{\rm sol}(x, z).
$
%
The scaled configuration $\vec A_\lambda(x,z)$ has the same topological charge as
\begin{eqnarray}
Q[\vec A_\lambda]&=&\frac{1}{2\pi} \int dx dz 
\left[\partial_x A_z^{\rm sol}(\lambda x,z) - \partial_z A_x^{\rm sol}(\lambda x,z)\right] 
\nonumber \\
&=&\frac{1}{2\pi} \int d\bar{x} dz 
\left[\partial_{\bar x} A_z^{\rm sol}(\bar{x},z) - \partial_z A_x^{\rm sol}(\bar{x},z)\right]
=Q[\vec A^{\rm sol}],
\end{eqnarray}
with $\bar{x}=\lambda x$.
The total energy $M$ of this scaled configuration $\vec A_\lambda(x,z) $ 
is $\lambda$ times of the original mass $M[\vec A^{\rm sol}]$: 
\begin{eqnarray}
M[\vec A_\lambda]
&=&\frac{N_c}{8\pi} \int dx dz f(z) 
\left[\partial_x A_z^{\rm sol}(\lambda x,z) - \partial_z \lambda A_x^{\rm sol}(\lambda x,z)\right]^2 \nonumber \\
&=&\lambda \frac{N_c}{8\pi} \int d\bar{x} dz f(z) 
\left[\partial_{\bar x} A_z^{\rm sol}(\bar{x},z) - \partial_z A_x^{\rm sol}(\bar{x},z)\right]^2
=\lambda M[\vec A^{\rm sol}].
\end{eqnarray}
Therefore, the total energy becomes smaller continuously to zero, 
when $\lambda$ goes to zero from unity. 
This leads to ``swelling instability'' of the topological configuration 
with any baryon number $B=Q$, as shown in Fig.~\ref{scale}.
%
Thus, in holographic QCD of 1+1 single-flavor QCD, 
all the (multi)baryonic configurations are unstable 
against this type of scale variation, 
and any topological (baryonic) configuration swells infinitely, 
at the leading of $1/N_c$ and $1/\lambda$ expansions.

\section{Comparison with Abrikosov Vortex in Type-II Superconductor}

\begin{figure}
\vspace{-7mm}
\begin{center}
\includegraphics[width=50mm,clip]{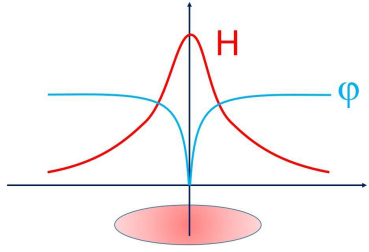}
\end{center}
\caption{The Abrikosov vortex in Type-II superconductors:
a stable topological configuration 
composed of the magnetic field $H$ and the Cooper-pair scalar field $\varphi$. 
The vortex has a topological charge on $\Pi_1(U(1))={\bf Z}$.}
\label{Abrikosov}
\end{figure}

Finally, to understand the physical reason of the swelling instability of 
the topological (baryonic) configuration in holographic QCD of 1+1 single-flavor QCD, 
we compare it with the Abrikosov vortex, which is a stable topological configuration 
appearing in the Type-II superconductor in an external magnetic field. 
(See Fig.~\ref{Abrikosov}.)

The superconducting theory consists of the photon field $A_\mu$ and 
the Cooper-pair scalar field $\varphi$, 
\begin{eqnarray}
{\cal H}= \frac{1}{2} \vec H^2+|(i\vec \partial+e\vec A)\varphi|^2 
+\lambda(|\varphi|^2-v^2)^2,
\end{eqnarray}
and has a topological charge of the Pontryagin index, 
e.g., $Q=\frac{1}{2\pi}\int dxdy H_z \in {\bf Z}$, 
corresponding to $\Pi_1(U(1)) \in {\bf Z}$, and 
the Abrikosov vortex is a topological configuration with $Q=1$. 

On the scale transformation, 
the photon-field contribution is to promote ``swelling'' of the soliton, 
and the scalar-field contribution is to promote ``shrinkage'' of the soliton.
Because of the competition between these two opposite effects, 
the Abrikosov vortex is stable against the scale transformation. 

On the other hand, holographic QCD of 1+1 QCD 
has only the vector field $A_M$ in 1+2 dimension, 
at the leading order of $1/N_c$ and $1/\lambda$.
On the scale transformation, the vector-field contribution is 
to promote swelling of the soliton.
Because of this one-side effect, 
the topological (baryonic) soliton is unstable against the scale transformation,
unlike the stable Abrikosov vortex in superconductors. 

\section{Summary}
We have studied baryons in holographic QCD corresponding to 1+1 dimensional 
single-flavor QCD for the first time. 
After formulating 1+1 QCD with an 
$S^1$-compactified D2/D8/$\overline{\rm D8}$ branes, 
we have described the baryon as a topological configuration in 1+1 $N_f$=1 QCD, 
corresponding to $\Pi_1({\rm U(1)})={\bf Z}$. 
Unlike 1+3 QCD with $N_f \ge 2$, 
we have found that the low-dimensional (multi)baryonic soliton is 
generally unstable against a scale transformation/variation and swells infinitely 
in 1+1 $N_f$=1 QCD at the leading of $1/N_c$ and $1/\lambda$. 
In this way, there is a serious difficulty on the soliton picture of baryons 
in large $N_c$ in the single-flavor world in both 1+1 QCD and 1+3 QCD.

\section*{Acknowledgements}
We thank Prof. S. Sugimoto for useful comments and discussions.

\end{document}